\documentclass{elsart}
\usepackage{natbib}
\usepackage{graphicx}
\begin{document}
\runauthor{Lair,et al.}
\begin{frontmatter}
\title{Late Light Curves of Normally-Luminous Type Ia Supernovae}
\author[Clemson University]{Jessica C. Lair}
\author[Clemson University]{Mark D. Leising}
\author[Steward Observatory]{Peter A. Milne}
\author[Steward Observatory]{G. Grant Williams}

\address[Clemson University]{Department of Physics and Astronomy, Clemson University, Clemson, SC 29634}
\address[Steward Observatory]{Steward Observatory, University of Arizona, Tucson, AZ 85721}
\begin{abstract}
                                                                                
The use of Type Ia supernovae as cosmological tools has reinforced the need to better understand these objects and their light curves.  The light curves of Type Ia supernovae are powered by the nuclear decay of $^{56}Ni \rightarrow ^{56}Co \rightarrow ^{56}Fe$.  The late time light curves can provide insight into the behavior of the decay products and their effect of the shape of the curves.  We present the optical light curves of six ``normal" Type Ia supernovae, obtained at late times with template image subtraction, and the fits of these light curves to supernova energy deposition models.

\end{abstract}
\end{frontmatter}

\section{Introduction}

Type Ia Supernovae (SNe Ia) are thought to be the thermonuclear explosion of a white dwarf \citep[see][and references therein]{2000A&ARv..10..179L}.  The light curves of SNe Ia are powered by deposition in the SN ejecta of the $\gamma$-ray and positron products of the $^{56}Ni\rightarrow ^{56}Co\rightarrow ^{56}Fe$ decay \citep*{1969ApJ...157..623C}. The extreme brightness and seemingly uniform light curves of SNe Ia make them good candidates for use as standard candle distance indicators. In more recent years, it has been shown that Type Ia supernovae do not have uniform light curve magnitudes, shape or spectra.  The light curves can, however, be normalized to account for this inhomogeneity, thus allowing these objects to be used at standard candle distance indicators \citep[e.g.][]{1993ApJ...413L.105P,1996ApJ...473...88R}.

Between 100-200 days after the explosion the ejecta become transparent to the $\gamma$-rays and the light curve is powered by the deposition of the positron kinetic energy into the ejecta.  The escape of a fraction of these positrons from the ejecta has been suggested as a possible source of the Galactic 511 keV annihilation radiation \citep*{1999ApJS..124..503M}. 

There are currently two methods of modeling the late emission of SNe Ia.  One is radiation transport with complete and instantaneous trapping of the positrons.  \cite{1980PhDT.........1A} showed, by comparing a model to the late time spectra of SN 1972E, that the ejecta will cool leading to an increased fraction of the emission coming out in the infrared, the so named ``infrared catastrophe" (IRC).  Other studies of radiation transport \citep[e.g.][]{1996ssr..conf..211F,2004A&A...428..555S}, have reproduced the IRC, but they also predict the abrupt fall off of the optical light curves as the emission shifts into the NIR and ultimately into the IR, which is not seen in observed light curves. 

The other method consists of positron energy deposition modeling without radiation transport.  In this type of modeling \citep[e.g.][]{1980ApJ...237L..81C,1997A&A...328..203C,1998ApJ...500..360R,1999ApJS..124..503M} , optical band light curves are used as tracers of the bolometric luminosity and fit to model energy deposition curves.  The results show model curves with varying degrees of positron escape fitting the light curves.  One weakness in this model fitting technique is in using the optical bands as tracers of bolometric.  \citet{2001ApJ...559.1019M} constructed bolometric curves using BVRI bands and showed those curves roughly fitting the positron escape energy deposition curves.

\section{BVRI Photometry using Template Subtraction}

We preformed aperture photometry on six ``normal" SNe Ia at late epochs, SN 2000E, SN 2000ce, SN 2000cx, SN 2001C, SN 2001bg, SN 2001dp.  Some of these SNe were located in very complicated regions in their host galaxies.  For this reason, we chose to do template image subtraction, on all but SN 2000cx, before preforming the aperture photometry.  All data reduction, image subtraction and aperture photometry was performed using the Image Reduction and Analysis Facility (IRAF) software \footnote{IRAF is distributed by the National Optical Astronomy Observatory.  http://iraf.noao.edu}.  The combined light curves can be seen in Figures \ref{radmodel} \& \ref{posmodel}, where they are normalized to be zero magnitude at 200days past explosion assuming an 18d rise time to peak light.  The data set for SN 2000E includes photometry from \cite{2003ApJ...595..779V}, and the data set for SN 2000cx includes data from \cite{2001PASP..113.1178L}, \cite{2002PhDT........10J}, \cite{2003PASP..115..277C}, and  \cite{2004A&A...428..555S}, where the data from our observations are plotted as the filled symbols.

\section{Light Curve Decline Rates}
\begin{figure}
\centerline{\includegraphics[scale=0.6]{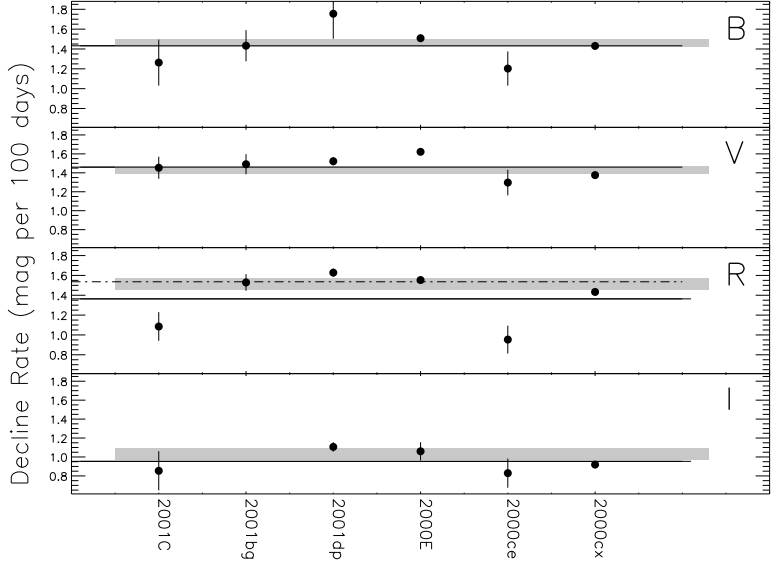}}
\caption{SN decline rates.\label{slopes}}
\end{figure}

The decline rates, the slope of the light curve, between 200-500 days were calculated for these light curves and the averages are shown in Figure \ref{slopes}, where the solid line is average for the six SNe. The calculated averages for B,V,R,\& I bands were 1.43 (0.07), 1.46 (0.04), 1.36(0.04), 0.95 (0.06), respectively, in magnitudes per day.  The shaded bar represents the average decline rate for 16 normal/super-luminous SNe Ia from \cite{2001ApJ...559.1019M} with a $1\sigma$ error bar.  In the R-band, there is a second average, represented by the dot-dashed line, which is the average decline rate leaving out SN 2000ce and SN 2001C.  This was done only to show the agreement with the Milne et al. averages.  

As shown in Figure \ref{slopes}, the B,V,\& R bands have decline rates of $\sim 1.4$ mag/day but the I-band has a much shallower slope of 0.95 mag/day.  This is in agreement with the decline rates of SN 2000cx as shown by \cite{2004A&A...428..555S}.  These results suggest that a slower I-band decline rate is a general feature of the late light curves of normal/super-luminous SNe Ia, and is possibly suggesting a shift in the late emission to longer wavelengths. A major result of \cite{2004A&A...428..555S} was the constant late time emission seen in the NIR curves of SN 2000cx, which supports the idea that the emission is moving into the NIR and eventually into the IR resulting in an IRC.  Our results from the analysis of these SNe reinforce the need for more observations of SNe Ia in the NIR in an attempt to reproduce what was seen in SN 2000cx and also in SN 1998bu \citep{2004A&A...426..547S}

\section{Discussion}

\begin{figure}
\centerline{\includegraphics[scale=0.6]{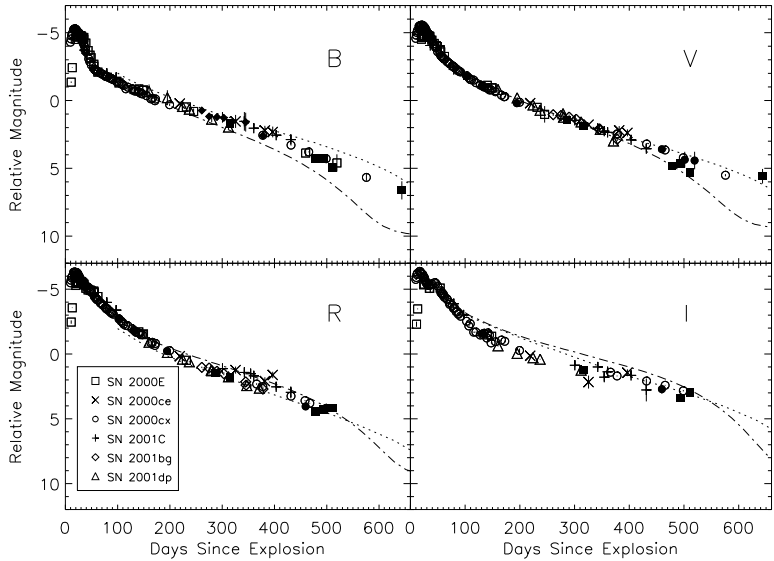}}
\caption{BVRI Light curves of 6 SNe Ia plotted on the radiation transport model light curves of \citet{2004A&A...428..555S}.\label{radmodel}}
\end{figure}

Figure \ref{radmodel} shows the combined light curves of the six SNe plotted on the radiation transport models of \cite{2004A&A...428..555S}.  The V-band model light curve has been normalized to be zero magnitude at 200d along with the data.  The B, R, \& I band model light curves have been adjusted so that the colors of the model are preserved.  The dotted curve is the model including photoionization and the dot-dashed curve is the model without photoionization.

\begin{figure}
\centerline{\includegraphics[scale=0.6]{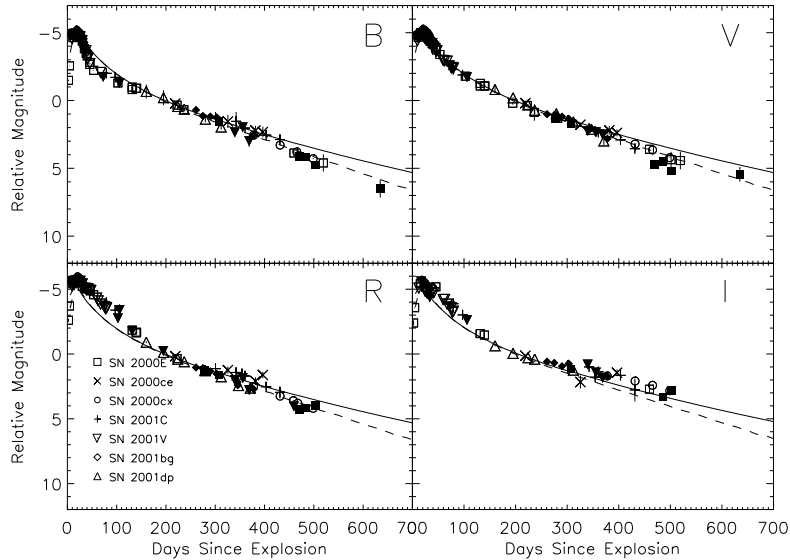}}
\caption{BVRI Light curves of 6 SNe Ia plotted on the postiron transport energy deposition curves of \citet{1999ApJS..124..503M}\label{posmodel}}
\end{figure}

Figure \ref{posmodel} shows the combined light curves of the six SNe plotted on the positron energy deposition curves of \cite{2001ApJ...559.1019M}.  The model curves have been normalized to be zero magnitude at 200d.  The solid curve is the energy deposition with the positron kinetic energy trapped and deposited into the ejecta and the dashed curve is the energy deposition curve with a radially combed magnetic field allowing a fraction of the positrons to escape the ejecta without depositing their kinetic energy.  

As can be seen in these figures, the shape of the B, V, \& R band light curves could be explained by either the color evolution in the radiation transport model or the escape of positrons from the ejecta, while the I-band has a slower decline rate than both models.  This suggests that a model combining radiation transport with positron transport would be preferred to attempt to explain the late light curves of SNe Ia.  One thing is clear from the light curves; these SNe show very little deviation from each other in a given band, implying that within this class of normally-luminous SNe Ia there is only one answer for the question of positron escape from SNe Ia ejecta.

\end{document}